\documentstyle[12pt,a4,epsf]{article}

\newcommand{\be}{\begin{equation}}
\newcommand{\ee}{\end{equation}}
\newcommand{\ba}{\begin{eqnarray}}
\newcommand{\ea}{\end{eqnarray}}
\newcommand{\dis}{\displaystyle}

\renewcommand{\mathrm}[1]{{\rm #1}}

\begin{document}
\begin{titlepage}
\begin{flushright}
{FTUV/98-84}\\{IFIC/98-85}\\{UG-FT-76/97}\\ {hep-ph/9811263}\\
\end{flushright}
\vspace{2cm}
\begin{center}
{\large\bf Strange Quark Mass Dependence \\ of the Tau Hadronic Width
\footnote{Invited talk at ``High Energy Conference on Quantum Chromodynamics
 (QCD '98)'', 2-8 July 1998, Montpellier. Work supported
in part by CICYT and DGESIC, Spain (Grants No. AEN-96/1672 and PB97-1261),
and by the European Union TMR Network $EURODAPHNE$ (Contract
No. ERBFMX-CT98-0169).}}\\
\vfill
{\bf Joaquim Prades $^a$ and Antonio Pich $^b$}\\[0.5cm]
$^a$ Departamento de
 F\'{\i}sica Te\'orica y del Cosmos, Universidad de Granada\\
Campus de Fuente Nueva, E-18002 Granada, Spain.\\[0.5cm]
$^b$  Departament de F\'{\i}sica Te\`orica, Universitat de Val\`encia and\\
 IFIC, CSIC - Universitat de Val\`encia,
 C/ del  Dr. Moliner 50, \\ E-46100 Burjassot (Val\`encia), Spain.\\[0.5cm]
\end{center}
\vfill
\begin{abstract}
\noindent
The perturbative quark mass corrections to the tau hadronic width
are studied 
to $O(\alpha_s^3 \, m_q^2)$. Including
up to dimension four corrections, we get $m_s(4 \, {\rm GeV}^2)=(143\pm42)$ MeV 
[$m_s(1\, {\rm GeV}^2)=(193\pm59)$ MeV ]. Possible improvements to reduce
the theoretical uncertainty are pointed out.
\end{abstract}
\vfill
September 1998
\end{titlepage}

\section{Introduction}\noindent 
The measurements of the tau hadronic width, 
\be
\label{defrtau}
R_\tau \equiv \frac{\dis \Gamma \left[
\tau^-\to {\rm hadrons} \, \nu_\tau (\gamma) \right]}
{\dis \Gamma \left[
\tau^- \to e^- \, \overline{\nu}_e \, \nu_\tau (\gamma) 
\right]} \, , 
\ee
and related observables 
have reached a high precision status, which will further improve
with the foreseen  data from heavy--flavour factories.
This will allow for much more detailed studies of the Standard Model
low--energy dynamics and in particular of QCD.

One of the important issues that can be addressed with these
measurements is the determination of the strange quark mass. 
The latest Review of Particle Physics (PDG) \cite{PDG98} quotes the 
($\overline{MS}$) running
strange quark mass at 2 GeV 
to be in between  60 MeV and 170 MeV. This large range reflects
mainly the uncertainty in the hadronic input needed to
determine $m_s$ within
QCD sum rules and  the spread of values
obtained within lattice QCD.

The strange quark mass induces a sizeable correction 
\cite{BNP:92} to the 
semi-inclusive $\tau$ decay width into Cabibbo--suppressed modes.
This can be used to perform a determination of $m_s$.
Preliminary values, extracted from the ALEPH $\tau$ data, have
been already reported in Refs. 
\cite{DAV97,CHE98,CDH98}. 
The obvious and very interesting advantage of this determination is
that the hadronic input does not depend on any extra hypothesis; it is
a purely experimental issue. Therefore, the major part of the uncertainty
will eventually come from the theoretical input, which can be treated
within QCD using the Operator Product Expansion (OPE)
at the tau mass scale.

It is then very important to have a detailed study of what do we
 know at present, within QCD, on the quark mass corrections to
$R_\tau$ and related observables.
This subject has been analyzed recently in 
Refs.~\cite{PP98a,PP98b,MAL98,KK:98,CKP98}.

\section{Theoretical Framework}\noindent
The theoretical framework needed to study the hadronic $\tau$ decay involves
two-point correlation functions of vector 
$V^\mu_{ij}\equiv \overline q_j \gamma^\mu q_i$  
and axial--vector quark currents
$A^\mu_{ij}\equiv\overline q_j \gamma^\mu \gamma_5 q_i$,
($i,j = u,d,s$):
\ba
\Pi^{\mu\nu}_{V, ij}(q) &\!\!\equiv&\!\! i {\dis \int }{\rm d}^4 x\, 
e^{iqx} \, \langle 0 | T \left\{ V_{ij}^{\mu\dagger}(x) V_{ij}^{\nu}(0) 
\right\} |0 \rangle \, ; \nonumber \\
\Pi^{\mu\nu}_{A, ij}(q) &\!\!\equiv&\!\! i {\dis \int }{\rm d}^4 x\, 
e^{iqx} \, \langle 0 | T \left\{ A_{ij}^{\mu\dagger}(x) A_{ij}^{\nu}(0) 
\right\} |0 \rangle \, ; \nonumber \\ 
\Pi^{\mu\nu}_{V(A), ij} &\!\!\equiv&\!\! 
\left(q^\mu q^\nu - q^2 \, g^{\mu \nu}\right)
\, \Pi^{L}_{V(A), ij}(q^2) \nonumber \\
&&\!\!\mbox{} + q^\mu q^\nu  \,\Pi^{T}_{V(A), ij}(q^2) . 
\ea
The tau hadronic width 
can be expressed as an integral over the 
invariant mass $s$ of the  final--state hadrons, of the spectral 
functions ${\rm Im} \, \Pi^T(s)$ and ${\rm Im} \, \Pi^L(s)$, 
with adequate  phase space factors:  
\ba
\label{rtau}
\lefteqn{R_\tau = 12 \pi {\dis \int^{M_\tau^2}_0} \frac{{\rm d} s}{M_\tau^2}
\, \left(1-{s\over M_\tau^2}\right)^2}&& \nonumber \\
&&\mbox{}\times \left[ 
\left( 1+2{s\over M_\tau^2}\right) {\rm Im}\, \Pi^T(s)
+ {\rm Im} \, \Pi^L(s) \right]\, . 
\ea
Moreover, according to the quantum numbers content of the
two--point function correlators   
\ba
\label{correlators}
\Pi^J(s) &\!\!\equiv&\!\!
 |V_{ud}|^2 \left[ \Pi_{V, ud}^J(s) + \Pi_{A,ud}^J(s)\right]
\nonumber \\ 
&\!\! +&\!\! |V_{us}|^2 \left[ \Pi_{V, us}^J(s) + \Pi_{A,us}^J(s)\right] ,
\ea
we can decompose $R_\tau$ into
\ba
R_\tau \equiv R_{\tau, V} + R_{\tau, A} + R_{\tau, S} \, ,
\ea
where $R_{\tau, V}$ and $R_{\tau, A}$ correspond to the first two terms in
Eq.~(\ref{correlators}), while $R_{\tau, S}$ contains the remaining
Cabibbo--suppressed contributions.

 Exploiting the analytic properties of $\Pi^J(s)$,
we can rewrite (\ref{rtau}) as a contour integral in the complex
$s$-plane  around the circle $|s|=M_\tau^2$ running counter--clockwise:
\ba
\label{contour}
\lefteqn{R_\tau = -i \pi \oint_{|s|=M_\tau^2} \, \frac{{\rm d}s}{s}
\left(1-{s\over M_\tau^2}\right)^3}&& \nonumber \\  &&
\times \left\{ 3  \left( 1 + {s\over M_\tau^2}\right)  D^{L+T}(s)
+ 4 D^L(s) \right\} \, .  
\ea
We used integration by parts to rewrite $R_\tau$ in terms of the logarithmic
derivative of the relevant correlators
\ba
D^{L+T}(s)&\!\!\equiv&\!\! -s \frac{{\rm d}}{{\rm d}s} \left[\Pi^{L+T}(s)\right]
\, , \nonumber \\
D^{L}(s)&\!\!\equiv&\!\! \frac{\dis s}{\dis M_\tau^2} 
\frac{\dis {\rm d}}{\dis {\rm d}s}  \left[s\, \Pi^{L}(s)\right] \, ,
\ea
which satisfy homogeneous renormalization group (RG) equations.

For large enough Euclidean $s$, $D^{L+T}(s)$ and $D^L(s)$
are calculable within QCD and we can organise the contributions 
in a series of higher dimensional contributions, using the OPE. One can then
express $R_\tau$ as an expansion in inverse powers of $M_\tau^2$
\cite{BNP:92}:
\ba
\label{deltas}
R_\tau \equiv  3 \left[ |V_{ud}|^2 + |V_{us}|^2 \right]
S_{\rm EW} \left[ \vphantom{{\dis \sum_{D=2,4,\cdots}}}
1 + \delta'_{\rm EW}\, + \delta^{(0)} 
\right. \nonumber \\ + \left. 
{\dis \sum_{D=2,4,\cdots}} \left( \cos^2(\theta_C) \, \delta_{ud}^{(D)}+
\sin^2(\theta_C) \, \delta_{us}^{(D)}  \right) \right] \, , \nonumber
\ea
where $\sin^2(\theta_C)\equiv |V_{us}|^2/[|V_{ud}|^2+|V_{us}|^2]$
and we have pulled out the electroweak
corrections  $S_{\rm EW}=1.0194\pm 0.0040 $ \cite{MS88} 
and $\delta'_{\rm EW}\simeq 0.0010$
\cite{BL90}. The dimension--zero contribution  $\delta^{(0)}$
is purely perturbative 
\cite{BNP:92,BR:88,NP:88,DP92}
and equal for the vector and axial--vector parts.
The symbols $\delta_{ij}^{(D)} \equiv [ \delta^{(D)}_{V, ij}
+ \delta^{(D)}_{A, ij}]/2$ are the average of the vector and axial--vector
contributions of the dimension $D\ge 2$ operators appearing in the
corresponding OPE.  

\section{Quark--Mass Corrections}\noindent
The largest contribution of the strange quark mass appears in 
$\delta_{us}^{(2)}$. This contribution was studied extensively 
in \cite{PP98a} and we shall report here the main results found.
Taking for simplicity $m_u=m_d=0$, the dimension two corrections
can be written as ($a\equiv \alpha_s/\pi$) 
\ba
\label{deltaus}
\delta^{(2)}_{us} &\!\!\equiv &\!\! -8 \, \frac{m_s^2(M_\tau^2)}{M_\tau^2}
\, \Delta \, [a(M_\tau^2)] \, , \\ \label{deltaus2}
\Delta[a]&\!\!\equiv&\!\! \frac{1}{4} 
 \left\{ 3\,\Delta^{L+T}[a(M_\tau^2)] +
 \Delta^{L}\, [a(M_\tau^2)]  \right\} , 
\ea 
where
\be
\label{delta2}
\Delta^{J}[a(M_\tau^2)] =  \sum_{n=0}\,  
\tilde  d_n^{J}(\xi) \,\, B^{(n)}_{J}(a_\xi) \, .
\ee
Here $\xi$ is an arbitrary scale factor (of order unity),
$a_\xi\equiv a(\xi^2 \, M_\tau^2)$ and the coefficients
$\tilde d_n^J(\xi)$ are constrained by the homogeneous 
RG equations satisfied by the corresponding 
$D^J(s)$. The question is how well can we predict $\Delta[a]$
within QCD.

The coefficients $\tilde d^J_n(\xi)$ are known
to $O(a^3)$ for $J=L$ and 
$O(a^2)$ for $J=L+T$.
The functions
$B^{(n)}_J(a_\xi)$ contain the contour integrations
\ba
\lefteqn{B^{(n)}_{L+T}(a_\xi) \equiv \frac{-1}{4 \pi i} \oint_{|x|=1}
\frac{{\rm d} x}{x^2} (1+x) (1-x)^3}&& \nonumber \\ &&\mbox{}\times
\left( \frac{m(-\xi^2 M_\tau^2 x)}{m(M_\tau^2)} \right)^2
a^n(-\xi^2 M_\tau^2 x) \, ;
\ea
and
\ba
\lefteqn{B^{(n)}_{L}(a_\xi) \equiv \frac{1}{2 \pi i} \oint_{|x|=1}
\frac{{\rm d} x}{x} (1-x)^3} && \nonumber \\  &&\mbox{}\times
\left( \frac{m(-\xi^2 M_\tau^2 x)}{m(M_\tau^2)} \right)^2
a^n(-\xi^2 M_\tau^2 x) \, . 
\ea
They have been calculated exactly \cite{PP98a}, using the RG to four loops; 
i.e. with the first four expansion coefficients
of the QCD beta and gamma functions.

The perturbative behaviour of $\Delta^{L+T}[a]$ 
and $\Delta^L[a]$ has been studied in Ref.~\cite{PP98a}.
For $\xi=1$ and $a=0.1$
(the actual value is around 0.11), one finds the following loop series:
\ba
\lefteqn{\Delta^{L+T}[0.1] =
0.7824+0.2239+0.0823} &&\nonumber \\ 
&& \mbox{} -0.000060 \, \tilde d^{L+T}_3(1) + \cdots
\ea
\ba
\lefteqn{\Delta^{L}[0.1] = 
1.5891+1.1733+1.1214}  &&\nonumber \\ 
&&\mbox{} + 1.2489 + \cdots \, .
\ea
While the $L+T$ series converges very well
[$\tilde d^{L+T}_3(1)$ is expected to be of $O(300)$],
the $L$ series behaves very badly. 
The combined final expansion for $\Delta$,
\ba
\lefteqn{\Delta[0.1] =
0.9840+0.4613+0.3421} &&\nonumber \\ 
&&\mbox{} +  [0.3122-0.000045 \,\tilde d^{L+T}_3(1)] + \cdots \, ,
\ea
looks still acceptable because $\Delta^{L+T}$ is weighted by a larger factor
in Eq.~(\ref{deltaus2}).

\begin{figure}[thb]
\begin{center}
\leavevmode\epsfxsize=7cm\epsfbox{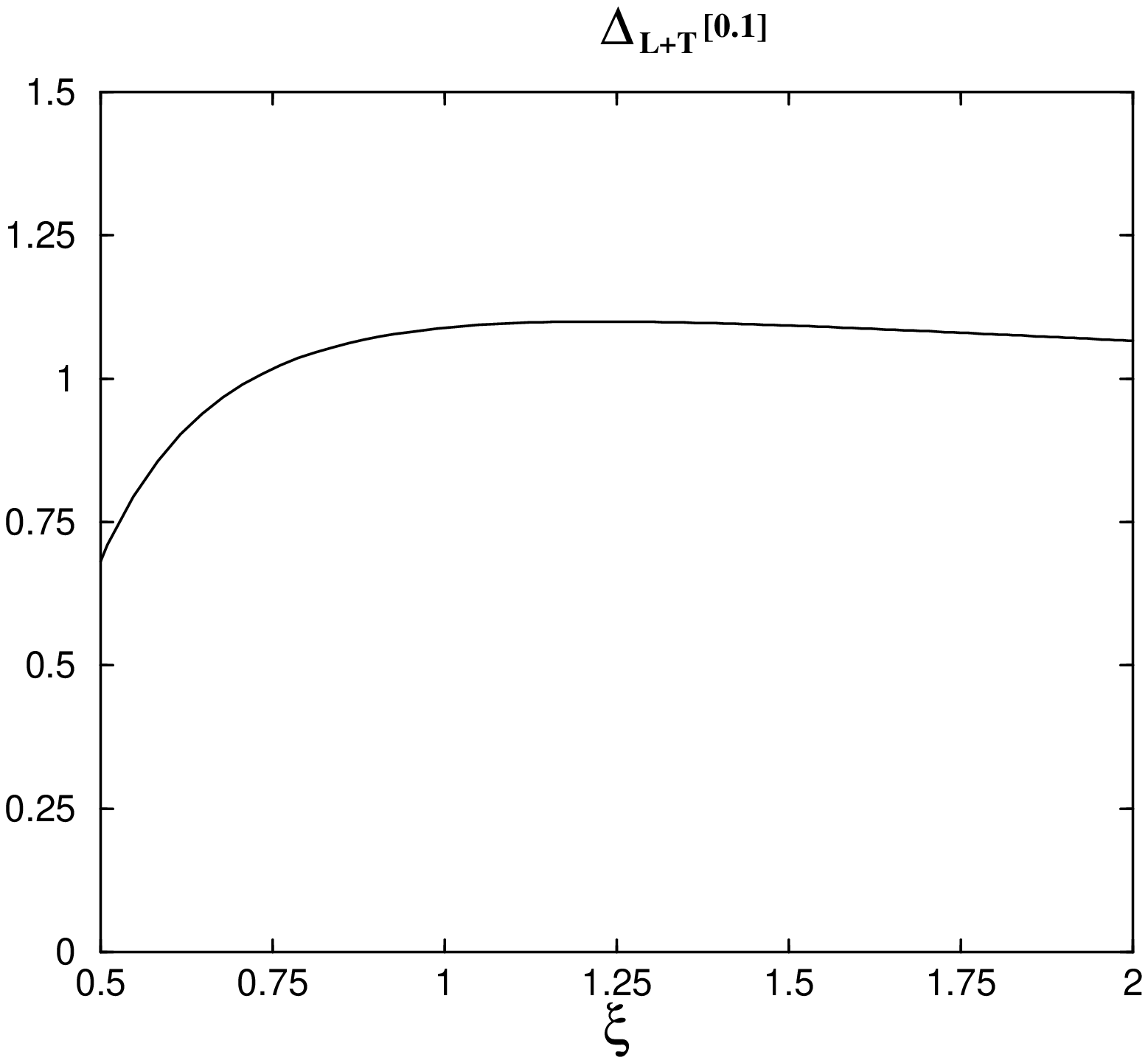}
\caption{\label{figDeltaLT} Variation of $\Delta^{L+T}[0.1]$
with the renormalization--scale  factor $\xi$, to four loops.}
\end{center}
\end{figure}
\begin{figure}[thb]
\begin{center}
\leavevmode\epsfxsize=7cm\epsfbox{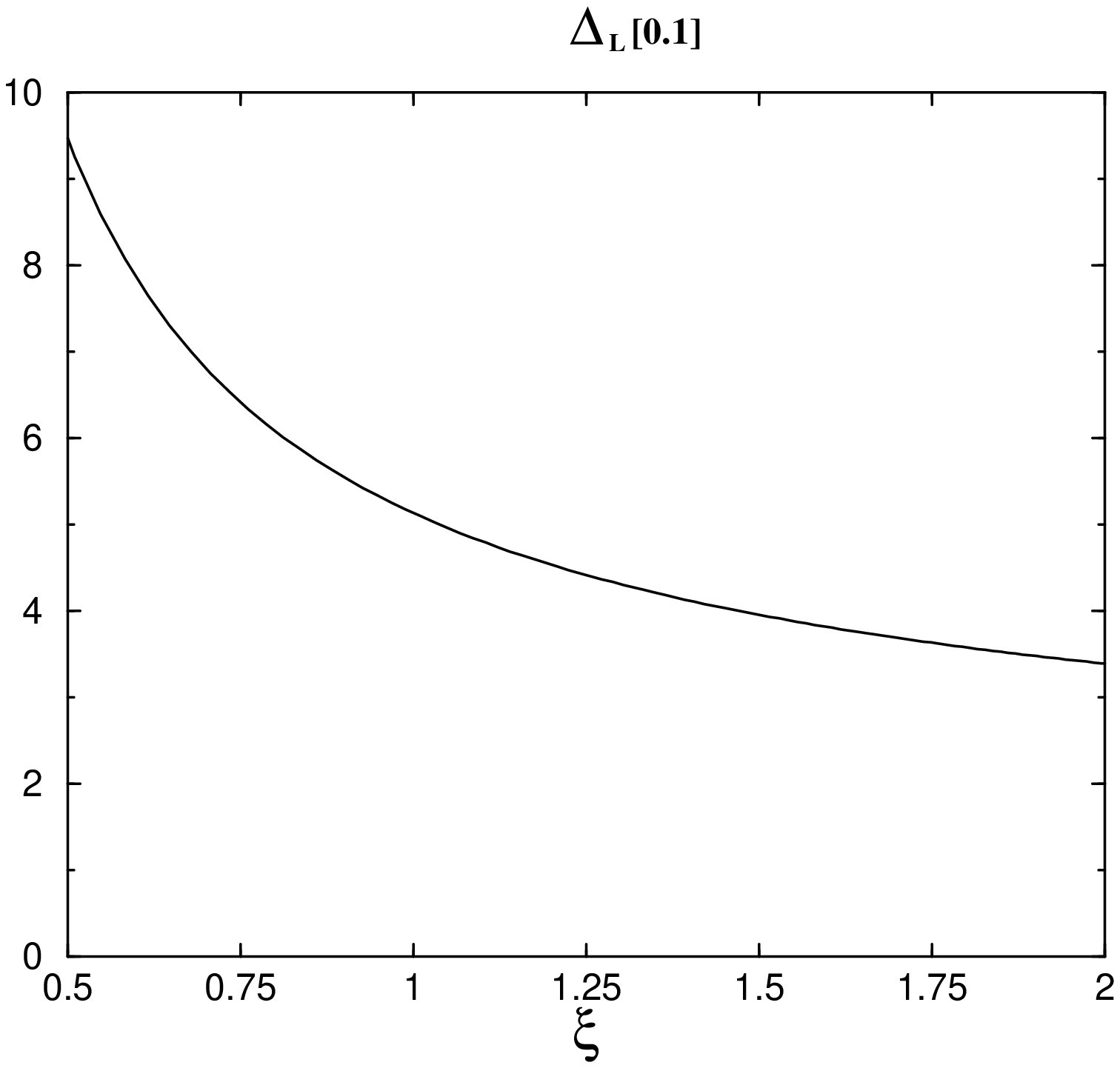}
\caption{\label{figDeltaL} Variation of $\Delta^L[0.1]$
with the renormalization--scale  factor $\xi$, to four loops.}
\end{center}
\end{figure}
\begin{figure}[thb]
\begin{center}
\leavevmode\epsfxsize=7cm\epsfbox{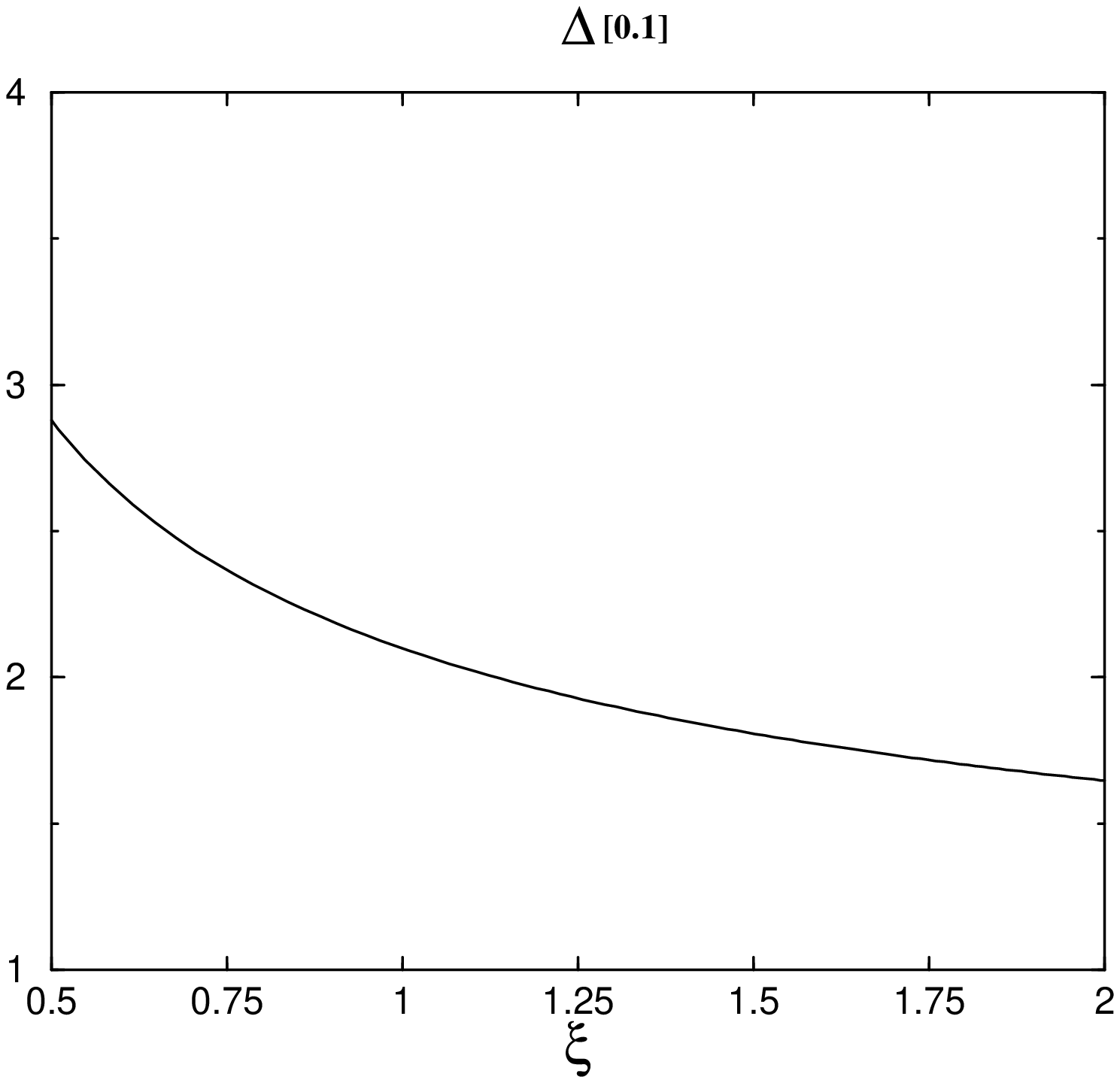}
\caption{\label{figDelta} Variation of $\Delta[0.1]$
with the renormalization--scale  factor $\xi$, to four loops.}
\end{center}
\end{figure}

The dependence on the renormalization--scale factor $\xi$ is shown \cite{PP98a}
in Figures~\ref{figDeltaLT} ($\Delta^{L+T}$),
\ref{figDeltaL} ($\Delta^{L}$) and 
\ref{figDelta} ($\Delta$).
The factor $\xi$ should be around one in order not to get large
logarithms. We have also to keep $a_\xi$ within  the radius of convergence
of the perturbative expansion \cite{DP92}, i.e. $\xi>0.5$.
Again, the $\Delta^{L+T}$ series behaves very well. In fact, one can
use some minimal sensitivity criterion to predict the unknown 
$\tilde d_3^{L+T}(1)$. On the contrary, the $\Delta^L$ series is 
monotonically decreasing and changes a 65\% between $\xi=1$ and 
$\xi=2$. Finally, $\Delta$ reflects  the bad 
$\Delta^{L}$ behaviour and is again monotonically decreasing.
Our best estimate of $\Delta[0.1]$ is \cite{PP98a}:
\be
\Delta[0.1] = 2.1\pm0.6 \, ,
\ee
where the central value is for $\xi=1$ and the error is a combination
of both the loop series expansion uncertainty and the truncation scale
dependence. 

\section{Results}\noindent
ALEPH has recently presented \cite{CDH98} a preliminary measurement of 
the Cabibbo--suppressed width of the $\tau$, 
$R_{\tau, S} = 0.1607 \pm 0.0066$. Moreover, they have extracted from their
data the SU(3)--breaking quantity,
\be\label{eq:ALEPHresult}
{{R_{\tau, V}+R_{\tau, A}} \over |V_{ud}|^2} - {R_{\tau, S}\over |V_{us}|^2}
= 0.413\pm 0.126 \, ,
\ee
which directly measures the effect of the strange quark mass.
The error includes the experimental uncertainty from $R_{\tau, S}$
as well as the one from 
the relevant quark mixing factors.

Including up to dimension four contributions,
Eq.~(\ref{eq:ALEPHresult}) implies \cite{PP98b}:
\ba
\label{mass}
m_s[M_\tau^2]&\!\! =&\!\! (149 \pm 44) \, \, {\rm MeV} , \nonumber \\
m_s[4\, {\rm GeV}^2]&\!\! =&\!\! (143 \pm 42) \, \, {\rm MeV} , \\
m_s[1\, {\rm GeV}^2]&\!\! =&\!\! (193 \pm 59) \, \, {\rm MeV} ,\nonumber
\ea
where the error, at the tau mass
scale, splits into 22 MeV from the experimental
uncertainty and 22 MeV from the theoretical one.
At present, the resulting error on $m_s$ is slightly larger than the 
usually quoted uncertainties from
QCD Sum Rules \cite{SR}
and lattice \cite{Lattice98} determinations.
Nevertheless, the $R_{\tau,S}$ result has the potential to be more precise
when better data will become available.
Notice that the lower value in (\ref{mass})
is already larger than the PDG  quoted lower bound
and excludes some of the lattice results \cite{Lattice98}.

The main theoretical uncertainty originates in the bad perturbative behaviour
of the longitudinal series $\Delta^L$. Therefore, an experimental
determination of the separate $J=L$ and $J=L+T$ pieces would allow a much more
precise analysis.

For the time being, let us
assume that we have just the full
final hadron mass distribution of $R_{\tau, S}$
(a first measurement of this distribution has already been presented in
Ref.~\cite{CDH98}). In that case
we could still reduce the theoretical uncertainty, through a
judicious choice of weight factors (i.e. moments) in Eqs.~(\ref{rtau})
and (\ref{contour}), which could improve the convergence
of the perturbative series (the phase space factors are partly responsible
for the bad perturbative behaviour of the $J=L$ contribution).
One could reach in this way a theoretical precision 
for the strange quark mass of the order or below
10 MeV
\cite{PP98b}. 
Obviously, a measurement  of the energy 
distribution $R_{\tau, S}(s)$ would also
decrease the experimental uncertainties considerably. 
We conclude then that there are good prospects 
for performing a precise
determination  of the strange quark mass from $\tau$ decays.

\end{document}